\documentclass[10pt,letterpaper]{article}
\usepackage{opex3}
\usepackage[draft]{hyperref}
\usepackage{bm}
\usepackage{amsmath}
\usepackage{amssymb}
\usepackage{graphicx,color}
\usepackage{epsfig}

\begin{document}

\title{Defocusing microscopy with an arbitrary size for the aperture of the objective lens}

\author{Ivan~F.~Santos,$^{1,*}$, W.~A.~T.~Nogueira,$^{2,3}$, S.~Etcheverry,$^{2,3}$, C.~Saavedra,$^{2,3}$, S.~P\'{a}dua$^{4}$ and G.~Lima,$^{2,3}$}
\address{$^1$ Departamento de F\'{i}sica, ICE, Universidade Federal de Juiz de Fora, \\ Juiz de Fora, CEP 36036-330, Brazil \\
$^2$  Center for Optics and Photonics, Universidad de Concepci\'{o}n, \\ Casilla 4016, Concepci\'{o}n, Chile \\
$^3$  Departamento de F\'{i}sica, Universidad de Concepci\'{o}n, \\  Casilla 160-C, Concepci\'{o}n, Chile \\
$^4$ Departamento de F\'{i}sica, Universidade Federal de Minas Gerais, \\ Caixa Postal 702, Belo Horizonte, Minas Gerais 30123-970, Brazil \\
$^*$Corresponding author: ivan@fisica.ufjf.br}

\begin{abstract}The theoretical approach to describe the defocusing microscopy technique by U. Agero et al. [Phys. Rev. E {\bf 67}, 051904 (2003)] assumes that the size of the objective lens aperture is infinite. This treatment gives that the intensity at the image plane depends on the laplacian of the phase introduced in the field by a pure phase object. In the present paper, we consider an arbitrary size for the aperture of the objective lens and we conclude that the intensity at the image plane depends also on the gradient of the phase introduced by the object and the phase itself. In this case, even an object that introduces only linear variations in the phase can be detected. Furthermore, we show that the contrast of the image of the phase object increases with the use of smaller objective apertures.
\end{abstract}


\section{Introduction}
Understanding how a defocused lens system affects a diffracted image has been the theme of several investigations \cite{Hopkins55,Faust55,Bryngdhal58,Stokseth69,Suzuki76,Stagaman,Nugent1,Nugent2,Agero03}. The experimental techniques and the developed theoretical background has been used as a tool for obtaining spatial and volumetric information of a microscopic phase object \cite{Suzuki76,Stagaman,Nugent1,Nugent2,Agero03,Ederlein03,tomo09,Nakahara11}. Throughout the last $60$ years many other microscopy techniques were developed for this purpose, such as the well-known Zernike phase-contrast method \cite{Zernike}, the digital speckle photography \cite{digital}, diffraction phase microscopy \cite{diffraction}, diffraction tomography \cite{tomography}, dark field microscopy \cite{Murphy}, differential interference contrast microscopy \cite{Murphy}, ghost images \cite{ghost} and  techniques which use the chromatic aberration \cite{aberration}. However, in comparison with these techniques, the defocusing microscopy technique (DM) is the simplest to implement, since we just need displace in the objective lens of the microscope from the image focusing distance.

Despite of its simplicity, DM is a very powerful technique, which has been applied in the research of biological systems. It was used to observe the shape, speed  and density of coherent propagating structures in macrophages (cell of the innate immune system), and measure the time of a single phagocytosis event \cite{Agero03}. Moreover, the shape, size, refractive index, bending modulus, and cytoplasm viscosity of red blood cells was obtained by applying DM to these cells \cite{RBC06}. DM was also used to do the tomography of fluctuating cell surfaces \cite{tomo09}. The curvature of the cell membrane is directly related to the laplacian of the phase introduced by it in the propagating field. On the other hand, the laplacian of the phase can be obtained from the contrast of the image observed when DM is used \cite{Agero03}. Thus, DM allows for obtaining the curvature of the surface of a cell directly, avoiding a subsequent calculation of the curvature.

Initially, the term ``Central illumination microscopy" was used to refer to DM \cite{Faust55}, where it was applied for determining the refractive index of transparent materials (refractometry). In this case, one usually observes an interference pattern which has been called ``Becke line" \cite{Barakat}. This pattern is symmetric when the phase edge varies abruptly and asymmetric when the phase varies slowly.

The theoretical approach for DM of Ref. \cite{Agero03} assumes that the size of the objective lens apertures is infinite. This approach predicts that the intensity at the image plane of a pure phase object depends only on the laplacian of the phase being modulated in the field. So, any linear phase change introduced by the object should not be detected by DM. In this paper we study the use of arbitrary apertures sizes for the objective lens and we show that by using small lenses, the intensity at the image plane depends also on the gradient of the phase and the phase itself. In this case, even an object that introduces only linear phase changes in the field can be detected. This improvement may lead to new applications for DM, such as the observation of ``hidden'' structures in cells.

Moreover, we study the dependence of the image contrast with the size of the objective lens aperture. We perform an experimental test of our theory and show that the image contrast increases when the entrance pupil of a microscope system is reduced. It is important to observe that for DM, to the best of our knowledge, there is no investigation about the role of the finite size of the lens aperture for the case of a coherent illumination. Only in the case of incoherent and partial coherent illuminations there have been studies showing that smaller lenses increases the contrast of the image due to the coherence introduced by spatial filtering \cite{Faust55,Stagaman}.

The rest of the article is structured as follows. In the following section, we make a short review of lens systems for imaging formation. In section $3$, we present the main theoretical results of this work, which is the extension of DM technique for the case of a finite objective lens aperture. In section $4$, we show a DM experiment in which it is possible to observe the effect of the finite size of the aperture in the contrast of the image of the phase object. It is exposed the concluding remarks of this manuscript in section $5$.

\section{Lens systems for image formation}

\subsection{The simplest lens system}
\begin{figure}[ht]
	\centering
		\includegraphics[width=0.8\textwidth]{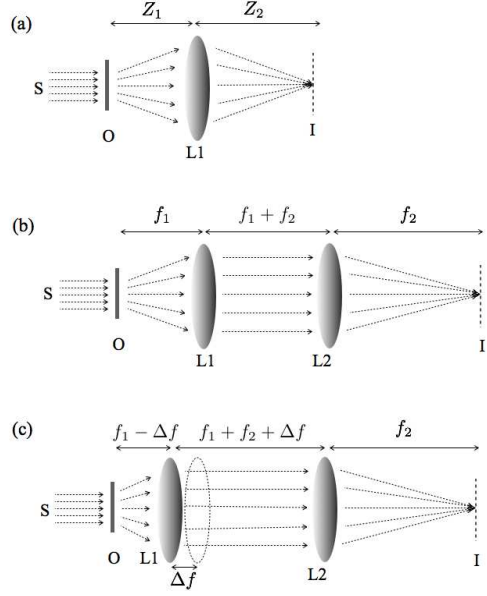}
		\caption{Schematic draw of three image systems. In these figures, S is a plane wave source that illuminates the object, O represents the object plane, I represents the image plane, and $f_{1}$ and $f_{2}$ are the focal lengths of the lenses L$1$ and L$2$, respectively. (a) Simplest system to observe the image of an ordinary object. Object and image planes are placed at distances $Z_{1}$ and $Z_{2}$, before and after the lens L$1$, respectively. (b) Microscope system. This system is used to generate the image of ordinary objects. The distance between lenses L$1$ and L$2$ is equal the sum of their respective focal lengths. (c) Defocusing microscopy scheme. $\Delta f$ is the length of the defocusing. This scheme is based on the one considered in Ref. \cite{Agero03}.}\label{fig1}
\end{figure}

The simplest imaging system uses a single lens, as shown in Fig. \ref{fig1}(a). S is a source of plane wave field. L$1$ is a lens, O is the object plane, and I is the image plane. In this case, the expression for the intensity of light at the position $\bm{\rho} =  x\overrightarrow{i} + y\overrightarrow{j}$ in the image plane, supposing a coherent illumination, is the square modulus of the following electric field \cite{Goodman}
\begin{eqnarray}
 E(\bm{\rho}) = A \times \int
d\bm{\xi} E_{0} \left( \bm{\xi} \right) T\left(\frac{k \bm{\xi}}{Z_{1}} + \frac{k \bm{\rho}}{Z_{2}} \right),
\label{eq ordinary}
\end{eqnarray}
where $Z_{1}$ is the distance from the object plane to the lens and $Z_{2}$ is the distance from the lens to the image plane. $k$ is the magnitude of the wave vector. $E_{0} \left( \bm{\xi} \right)$ is the light electric field at the position $\bm{\xi} = \xi_{x}\overrightarrow{i} + \xi_{y}\overrightarrow{j}$ immediately after the object plane. $T\left(\frac{k \bm{\xi}}{Z_{1}} + \frac{k \bm{\rho}}{Z_{2}} \right)$ is the Fourier transform of the transmission function of the lens aperture. The term $A$ before the integral in Eq. (\ref{eq ordinary}) contains only scale factors and phase components that do not affect the form of the image intensity. All variables which appear in boldface in all equations of this paper mean vectors in the transverse plane having $x$ and $y$ components. The Eq. (\ref{eq ordinary}) is valid whether one of the following two conditions are satisfied: (i) the object is illuminated  by a spherical wave that is converging towards the point where the optical axis pierces the lens \cite{Goodman} or (ii) the size of object is no greater than about $1/4$ the size of the lens aperture \cite{Goodman,Tichenor}.

\subsection{A simple microscope system}
A slightly more complex imaging system discussed in  \cite{Agero03} is the microscope constituted by two lenses, as can be seen in Fig. \ref{fig1}(b). Considering a finite aperture for the objective lens and an infinite aperture for the eyepiece lens, the expression for the intensity of light in a position $\bm{\rho}$ at the image plane, supposing a coherent illumination, is given by (see appendix)
\begin{eqnarray}
 E(\bm{\rho}) = B \times \int
d\bm{\xi} E_{0} \left( \bm{\xi} \right)  T_{1}\left(\frac{k \bm{\xi}}{f_{1}} + \frac{k \bm{\rho}}{f_{2}} \right),
\label{eq telescopic}
\end{eqnarray}
where $f_{1}$  and $f_{2}$ are the focal lengths of the lenses considered in the setup. $T_{1}\left(\frac{k \bm{\xi}}{f_{1}} + \frac{k \bm{\rho}}{f_{2}} \right)$ is the Fourier transform of the transmission function of the objective lens aperture. $B$ is a term that includes only scale factors and phases. Since we are interested in the intensity detected at the image plane, from now on we will no longer write this term in order to simplify the mathematical manipulation. Note that to form an image, it is assumed for Eq. (\ref{eq telescopic}) the same conditions considered for Eq. (\ref{eq ordinary}) mentioned in the last section \cite{Goodman,Tichenor}.

\subsection{Ideal imaging systems}
Better image resolutions can be obtained by lenses with large apertures. By assuming infinite apertures in both Eqs. (\ref{eq ordinary}) and (\ref{eq telescopic}), their Fourier transforms become delta functions which filter the integrals. In the first case [Eq. (\ref{eq ordinary})] we obtain
\begin{eqnarray}
E(\bm{\rho})  &=&   E_{0} \left(   -\frac{{Z_{1}}}{Z_{2}} \bm{\rho}  \right).
\label{eq perfect ordinary}
\end{eqnarray}
In the second case [Eq. (\ref{eq telescopic})] we obtain
\begin{eqnarray}
E(\bm{\rho})  &=&  E_{0} \left(  - \frac{ {f_{1}}}{f_{2}} \bm{\rho}  \right).
\label{eq perfect telescopic}
\end{eqnarray}
In both cases, the minus sign indicates that the image is inverted when compared to the object.

\subsection{Image formation of pure phase objects without special methods}
For a pure phase object and uniform illumination, we have $ E_{0} \left( \bm{\xi} \right) = e^{i\phi\left( \bm{\xi} \right)}$, which results that the square modulus of both Eqs. (\ref{eq perfect ordinary}) and (\ref{eq perfect telescopic}) are uniform for infinite aperture lenses, since this last condition means that $T$ and $T_{1}$ are Dirac delta functions. In the case of an amplitude object, a decrease of the size of the lens aperture implies in a decrease of the resolution of the image \cite{Klein,Ivan03}. So, one could expect that the use of lenses with smaller apertures would make the methods described in sections $3$ and $4$ even more useless for the image formation of pure phase objects. Fortunately, this is an incorrect assumption (see Ref. \cite{Barakat}). In this section we discuss an example in order to show this.

\begin{figure}[ht]
\centering
\includegraphics[width=0.7\textwidth]{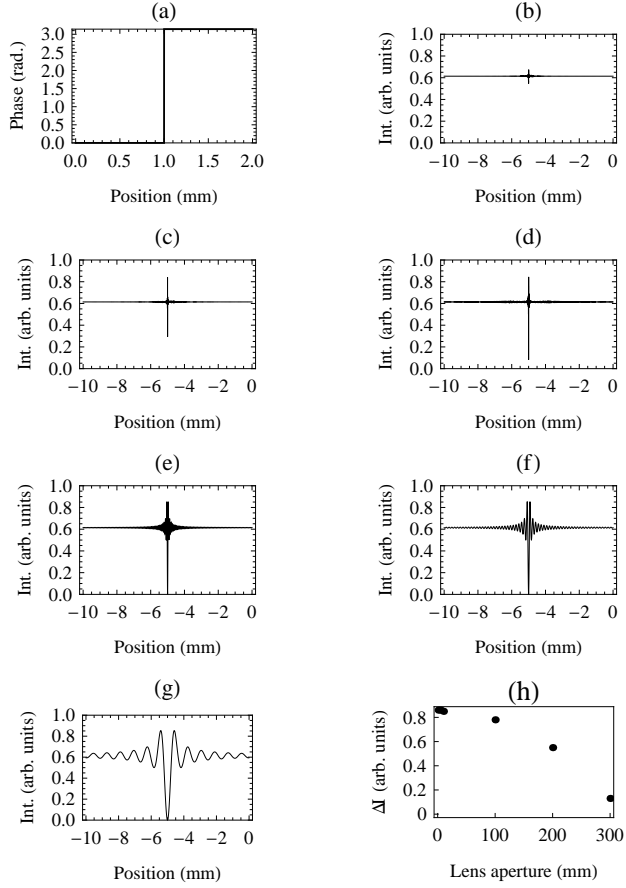}
\caption{(a) Phase introduced by a pure phase object. It has a discontinuity at the position $1$\,mm. (b)-(g) Intensity of the images produced by cylindrical lenses with different widths. The intensities, which were obtained assuming an ordinary microscope system (i.e. without defocusing), were calculated by using Eq. (\ref{eq telescopic}). The parameters used in the simulations are $f_{1}= 100$\,mm, $f_{2}=500$\,mm, and $826$\,nm of wavelength. The widths were $2L=300$\,mm, $2L=200$\,mm, $2L=100$\,mm, $2L=10$\,mm, $2L=5$\,mm and $2L=1$\,mm in figures (b), (c), (d), (e), (f) and (g), respectively. (h) The maximum variation of the intensity ($\Delta I_{max} = I_{max} - I_{min}$) in figures (b)-(g) as a function of the width $2L$.}
\label{sem}
\end{figure}

Let us consider a pure phase object that introduces a $\pi$ of discontinuity in the phase of the incident field. This discontinuity is located at the position $1$\,mm from the optical axis of propagation of the field at the plane of the object, as can be seen in Fig. \ref{sem}(a). Figures \ref{sem}(b)-(g) show  the intensity of the light at the image plane. These intensities are calculated by using Eq. (\ref{eq telescopic}) and considering the widths of the cylindrical lenses equal to $2L=300$\,mm, $2L=200$\,mm, $2L=100$\,mm, $2L=10$\,mm, $2L=5$\,mm and $2L=1$\,mm, respectively. For simplicity, we consider an one-dimensional case. We suppose the use of a bi-concave cylindrical lens, that provides magnification only in one dimension. This implies that the one-dimensional Fourier transform in Eq. (\ref{eq telescopic}) is given by $sinc\left(\frac{kL \xi}{f_{1}} + \frac{kL \rho}{f_{2}} \right)$.

Note that the fringes around the position $-5$\,mm  are owing to the phase discontinuity in the position $1$\,mm at the object plane. These positions are related by the magnification $M=-f_{2}/f_{1}=-5$. It is important to note that due to the small size of the objective it is possible to detect the discontinuity by means of the usual method of image formation, without any defocusing or introduction of some other contrast technique. That is a result that can not be obtained if infinite lenses are considered. In Fig. \ref{sem} we can also note that the diffraction bands at the image plane become narrower when the size of the objective lens increases, which is in agreement with Ref. \cite{Faust55}.

Figure \ref{sem}(h) shows the maximum variation of the intensity in the image plane with respect to different widths of the cylindrical lenses. This variation is defined as $\Delta I_{max} = I_{max} - I_{min}$ and it is directly related to the contrast of the image \cite{Agero03}. Figure \ref{sem}(h) shows that objectives with small apertures (i.e. smaller values of $L$) generates images with higher contrasts. This behavior means that a restriction in the linear dimension of the lens aperture improves the capability to detect phase objects.

\section{Defocusing microscopy}

\subsection{Defocusing microscopy with an objective lens of arbitrary size aperture}

The DM consists of making a small displacement at the objective longitudinal position, as shown in Fig. \ref{fig1}(c). This displacement allows the detection and reconstruction of phase objects. The treatment of Ref. \cite{Agero03} assumes that the lenses apertures are infinite. This approach gives the following expression for the intensity at the image plane
\begin{eqnarray}
I(\bm{\rho})= 1 + \frac{\Delta f}{k}\nabla_{\rho'}^2 \left[ \phi\left( \bm{\rho}'  \right)\right],
\label{eq bira}
\end{eqnarray}
where $\bm{\rho}' = -\frac{f_{1}}{f_{2}} \bm{\rho}$ and the subscript $\rho'$ indicates that the laplacian is calculated with respect to this variable. The first term of Eq. (\ref{eq bira}) is associated to the backlight, which is always present in the image formation of phase objects. The parameter  $\Delta f$ in the second term is the amount of defocusing [see Fig.  \ref{fig1}(c)]. It is important to observe that any linear change in the phase is not taken into account by Eq. (\ref{eq bira}), owing to the fact that the laplacian of a linear function is zero.

\subsection{Defocusing microscopy with an arbitrary size for the objective lens aperture}
In this section we will use the Fresnel propagator as a tool in order to study the case of DM with finite size for the objective lens aperture. Consider the microscope system depicted in Fig. \ref{fig1}(c). The electric field at the image plane is given by
\begin{eqnarray}
E(\bm{\rho}) &=& \int
d\bm{\xi} E_{0} \left( \bm{\xi} \right)   \int d\bm{\alpha} \left| A_{1}(\bm{\alpha})\right | e^{\frac{-ik \alpha ^{2}}{2f_{1}}}    e^{\frac{ik |\bm{\alpha} - \bm{\xi|}^{2}}{2(f_{1} - \Delta f)}}   \nonumber         \\      &&  \times  \int d\bm{\beta}\left| A_{2}(\bm{\beta}) \right | e^{\frac{-ik \beta ^{2}}{2f_{2}}}    e^{\frac{ik |\bm{\beta} - \bm{\alpha|}^{2}}{2(d+\Delta f)}}    e^{\frac{ik |\bm{\rho} - \bm{\beta|}^{2}}{2f_{2}}} ,
\label{a}
\end{eqnarray}
where $E_{0} \left( \bm{\xi} \right)$ is the electric field immediately after the object plane. $\left| A_{1}(\bm{\alpha})\right| $ and $\left| A_{2}(\bm{\beta})\right| $ are the modulus of the transmission function of the objective and eyepiece, respectively. $f_{1}$ and $f_{2}$ are the focal lengths of the objective and eyepiece lenses, respectively. The integration in the variables $\bm{\xi}$,  $\bm{\alpha}$ and  $\bm{\beta}$ are due to the propagation from the object to the objective, objective to the eyepiece and eyepiece to the image, respectively. Eq. (\ref{a}) can be written as
\begin{eqnarray}
E(\bm{\rho}) =   e^{\frac{ik \rho ^{2}}{2f_{2}}}   \int d\bm{\xi}    E_{0} \left( \bm{\xi} \right)   \int d\bm{\alpha}
\left| A_{1}(\bm{\alpha})\right |
e^{\frac{ik |\bm{\alpha} - \bm{\xi|}^{2}}{2(f_{1} - \Delta f)}}  e^{\frac{-ik \alpha ^{2}}{2f_{1}}}  e^{\frac{ik \alpha ^{2}}{2(d+\Delta f)}}   h(\bm{\alpha} , \bm{\rho}),
\label{campo}
\end{eqnarray}
where
\begin{eqnarray}
h(\bm{\alpha} , \bm{\rho}) =  \int d\bm{\beta}  e^{\frac{-ik \beta ^{2}}{2f_{2}}}   e^{\frac{ik [\beta ^{2} -2 \bm{\beta}\cdot \bm{\alpha}]}{2(d+\Delta f)}}    e^{\frac{ik [\beta ^{2} -2 \bm{\beta}\cdot \bm{\rho}]}{2f_{2}}}. \label{yy}
\end{eqnarray}
In this last step we assumed that the eyepiece is sufficiently large so that $\left| A_{2}(\bm{\beta})\right| \approx 1$. The exponentials depending on  $\bm{\beta}^2$ cancel out, and the expression (\ref{yy}) simplifies to
\begin{eqnarray}
h(\bm{\alpha} , \bm{\rho}) =  C \times \  e^{\frac{-ik}{2(d+\Delta f)}   [\bm{\alpha} + \frac{(d+\Delta f)}{f_{2}} \bm{\rho} ]^2},
\label{h de alfa ro}
\end{eqnarray}
where $C$ is a term that represents only scale factors that does not depend upon the variables $\bm{\alpha}$ and $\bm{\rho}$ and will be dropped out. Inserting Eq. (\ref{h de alfa ro}) into Eq. (\ref{campo}), expanding all quadratic exponentials, and performing the algebraic sum of quadratic terms in the variable $\bm{\alpha}$, we obtain that
\begin{eqnarray}
E(\bm{\rho}) &=&  e^{\frac{ik}{2}\left[ \frac{1}{f_{2}} - \frac{f_{1}+f_{2}+\Delta f}{f_{2}^2}  \right] \bm{\rho}^{2}}
\int d\bm{\xi} E_{0} \left( \bm{\xi} \right) e^{\frac{ik}{2(f_{1} - \Delta f)}\bm{\xi}^{2}}  \nonumber \\  && \times
\int d\bm{\alpha} \left| A_{1}(\bm{\alpha})\right| e^{\frac{ik}{2}    \left( \frac{1}{f_{1} - \Delta f }   -  \frac{1}{f_{1}} \right)     \bm{\alpha}^{2}} e^{-ik \left( \frac{\bm{\xi}}{f_{1} - \Delta f }   +  \frac{\bm{\rho}}{f_{2}} \right)   \cdot  \bm{\alpha}}.
\label{ClassDefocGeral}
\end{eqnarray}
Note that Eq. (\ref{ClassDefocGeral}) is valid even for large values of $\Delta f$. Now let us suppose that $\Delta f$ is small, so that we can write  $\frac{1}{f_{1} - \Delta f } \approx \frac{1}{f_{1}} + \frac{\Delta f}{f_{1}^2}$ (this expression is obtained by Taylor expansion around $\Delta f = 0$). Thus, the first exponential expression in the second integral of Eq. (\ref{ClassDefocGeral}) becomes
\begin{eqnarray}
e^{\frac{ik}{2} \left( \frac{1}{f_{1} - \Delta f }   -  \frac{1}{f_{1}} \right)     \bm{\alpha}^{2}} \approx 1 + \frac{ik \Delta f}{2f_{1}^2}\bm{\alpha}^{2}.
\label{}
\end{eqnarray}
Unless a phase factor (which is irrelevant to the calculation of the intensity), the eletric field is given by
\begin{eqnarray}
E(\bm{\rho}) =  \int d\bm{\xi} E_{0} \left( \bm{\xi} \right) e^{\frac{ik}{2(f_{1} - \Delta f)}\bm{\xi}^{2}}
\int d\bm{\alpha} \left| A_{1}(\bm{\alpha})\right|    \left(  1 + \frac{ik \Delta f}{2f_{1}^2}\bm{\alpha}^{2}   \right)       e^{-ik \left( \frac{\bm{\xi}}{f_{1} - \Delta f }   +  \frac{\bm{\rho}}{f_{2}} \right)   \cdot  \bm{\alpha}}.
\label{}
\end{eqnarray}
By using the identity
\begin{eqnarray}
\nabla^2 \left[  e^{-ik \left( \frac{\bm{\xi}}{f_{1} - \Delta f }   +  \frac{\bm{\rho}}{f_{2}} \right)   \cdot  \bm{\alpha}} \right] =          \frac{-k^2 \bm{\alpha}^2}{f_{2}^2} e^{-ik \left( \frac{\bm{\xi}}{f_{1} - \Delta f }   +  \frac{\bm{\rho}}{f_{2}} \right)   \cdot  \bm{\alpha}},
\end{eqnarray}
it is possible to show that
\begin{eqnarray}
E(\bm{\rho}) &=&  \int d\bm{\xi} E_{0} \left( \bm{\xi} \right) e^{\frac{ik}{2(f_{1} - \Delta f)}\bm{\xi}^{2}} T_{1}\left(\frac{k \bm{\xi}}{f_{1} - \Delta f} + \frac{k \bm{\rho}}{f_{2}} \right) \nonumber \\  &&
-\frac{i \Delta f f_{2}^2}{2kf_{1}^2}\nabla^2 \int d\bm{\xi} E_{0} \left( \bm{\xi} \right) e^{\frac{ik}{2(f_{1} - \Delta f)}\bm{\xi}^{2}} T_{1}\left(\frac{k \bm{\xi}}{f_{1} - \Delta f} + \frac{k \bm{\rho}}{f_{2}} \right),
\label{Gen_field}
\end{eqnarray}
where $T_{1}$ is the Fourier transform of $\left| A_{1}(\bm{\alpha})\right|$. Now let us define the function $E_{0}^\prime(\bm{\rho})$ as
\begin{eqnarray}
E_{0}^\prime(\bm{\rho}) =  \int d\bm{\xi} E_{0} \left( \bm{\xi} \right) e^{\frac{ik}{2(f_{1} - \Delta f)}\bm{\xi}^{2}} T_{1}\left(\frac{k \bm{\xi}}{f_{1} - \Delta f} + \frac{k \bm{\rho}}{f_{2}} \right).
\label{def_field}
\end{eqnarray}
Note that $|E_{0}^\prime(\bm{\rho})|^2 $ give us the image of the phase object generated by a microscope system whose magnification is $ M= \frac{f_{2}}{f_{1} - \Delta f}  \approx \frac{f_{2}}{f_{1}} $. It is valid for a coherent illumination or for a partial coherent illumination when the object is smaller than the transverse coherence area of the field, as pointed out in Ref. \cite{Agero03}. From now on, it will be assumed that one of these two physical conditions is respected. By using the above definition in Eq. (\ref{Gen_field}), we have
\begin{eqnarray}
E(\bm{\rho}) = E_{0}^\prime(\bm{\rho}) \ \ - \frac{i \Delta f f_{2}^2}{2kf_{1}^2}\nabla^2 E_{0}^\prime(\bm{\rho}).
\label{}
\end{eqnarray}
This expression contains the laplacian of the Fourier transform of the modulus of the objective lens transmission function, the only part dependent on $\bm{\rho}$.  Let us try to get an expression containing the laplacian of the object function [as in Eq. (\ref{eq bira})]. By remembering that
\begin{eqnarray}
T_{1}\left(\frac{k \bm{\xi}}{f_{1} - \Delta f} + \frac{k \bm{\rho}}{f_{2}} \right) = \int d\bm{u} |A_{1}\left( \bm{u}\right)| e^{-i\left(\frac{k \bm{\xi}}{f_{1} - \Delta f} + \frac{k \bm{\rho}}{f_{2}} \right) \cdot \bm{u}},
\end{eqnarray}
it is possible to show that
\begin{eqnarray}
E(\bm{\rho}) &=&  E_{0}^\prime(\bm{\rho}) \ \ - \frac{i \Delta f f_{2}^2}{2kf_{1}^2}  \ \ \frac{(f_{1} - \Delta f)}{k}   \int d\bm{v}  T_{1}(\bm{v})         \nonumber \\  &&  \times   \nabla^2 \left[
E_{0} \left(  (f_{1} - \Delta f) ( \frac{\bm{v} }{k}   -   \frac{\bm{\rho} }{f_{2}})  \right)  e^{\frac{ik}{2}   (f_{1} - \Delta f) ( \frac{\bm{v} }{k}   - \frac{\bm{\rho} }{f_{2}})^2 }  \right].
\label{}
\end{eqnarray}
This expression contains the laplacian of the function of the object, however $E_{0}$ is calculated at the coordinate  $(f_{1} - \Delta f) (\frac{\bm{v} }{k}  -  \frac{\bm{\rho} }{f_{2}})$ instead of the coordinate  $\bm{v} $.  On the other hand, the Fourier transform which is usually calculated at the coordinate $( \frac{\bm{v} }{f_{1}}   +   \frac{\bm{\rho} }{f_{2}})$ [see Eq. (\ref{eq telescopic})], here is evaluated at $\bm{v}$. In order to write the above expression in a simpler form, let us make the change of variable given by
\begin{eqnarray}
(f_{1} - \Delta f) ( \frac{\bm{v} }{k}   -   \frac{\bm{\rho} }{f_{2}}) =  \bm{\xi}.
\end{eqnarray}
This change implies that $\bm{v} = \frac{k\bm{\xi} }{f_{1} - \Delta f} + \frac{k\bm{\rho} }{f_{2}}, \ \ $      $d\bm{v} = \frac{k }{f_{1} - \Delta f}  d\bm{\xi} \ \ $ and $\nabla^2 \equiv \nabla^2_{\bm{\rho}} = \left(\frac{f_{1} - \Delta f}{f_{2}}\right)^2 \nabla^2_{\bm{\xi}}$, where the subscripts indicate the variables for which the laplacian is calculated.
Using this change of variable and neglecting the terms that contain powers of  $\Delta f$  larger than or equal $2$ (as it was done in \cite{Agero03}), we obtain that
\begin{eqnarray}
E(\bm{\rho}) =  \int d\bm{\xi}  \left[  E_{0}(\bm{\xi})e^{\frac{ik\xi ^2}{2(f_{1}- \Delta f)}}  - \frac{i\Delta f}{2k}\nabla^2_{\bm{\xi}}\left( E_{0}(\bm{\xi})e^{\frac{ik\xi ^2}{2(f_{1}- \Delta f)}}  \right)   \right]  T_{1}\left( \frac{k\bm{\xi} }{f_{1} - \Delta f}   +   \frac{k\bm{\rho} }{f_{2}}\right).
\label{camp defoc obj qq}
\end{eqnarray}
Eq. (\ref{camp defoc obj qq}) is valid for small defocusing and any general class of objects. Let us now suppose a pure phase object illuminated by a plane wave with amplitude $E_{0i}$, such that the transmitted complex field amplitude is given by
\begin{eqnarray}
E_{0}(\bm{\xi}) = E_{0i}e^{i\phi (\bm{\xi})} .
\label{}
\end{eqnarray}
We also assume that the size of the object is smaller than about $1/4$ of the size of the lens aperture, so that we can neglect the quadratic phase factors $e^{\frac{ik\xi ^2}{2(f_{1}- \Delta f)}}$ according to \cite{Goodman,Tichenor}.
In this way, it follows that the electric field is given by
\begin{eqnarray}
E(\bm{\rho}) =  \int d\bm{\xi} E_{0i} \left[ e^{i\phi(\bm{\xi})}  - \frac{i\Delta f}{2k}\nabla^2_{\bm{\xi}}\left( e^{i\phi(\bm{\xi})} \right)   \right]  T_{1}\left( \frac{k\bm{\xi} }{f_{1} - \Delta f}   +   \frac{k\bm{\rho} }{f_{2}}\right).
\label{}
\end{eqnarray}
By using again the identity
\begin{eqnarray}
\nabla^{2}\left(e^{i\phi} \right) = e^{i\phi} \left[ -(\nabla \phi)^2 +i\nabla^{2}\phi \right]
\label{},
\end{eqnarray}
we obtain
\begin{eqnarray}
E(\bm{\rho}) =  \int d\bm{\xi} E_{0i} e^{i\phi(\bm{\xi})}  \left[1 +   \frac{\Delta f}{2k} \nabla_{\bm{\xi}}^2  \phi(\bm{\xi})      + \frac{i\Delta f}{2k}\left( \nabla_{\bm{\xi}}  \phi(\bm{\xi}) \right)^2   \right]                            T_{1}\left( \frac{k\bm{\xi} }{f_{1} - \Delta f} + \frac{k\bm{\rho} }{f_{2}}\right).
\label{camp class obj fase}
\end{eqnarray}
This is the final expression for the electric field. The intensity at the image plane, $I(\bm{\rho})=\left| E(\bm{\rho}) \right|^2$, is given by
\begin{eqnarray}
I(\bm{\rho}) &=& \int d\bm{\xi} E_{0i} e^{i\phi(\bm{\xi})}  \left[1 +   \frac{\Delta f}{2k} \nabla_{\bm{\xi}}^2  \phi(\bm{\xi})      + \frac{i\Delta f}{2k}\left( \nabla_{\bm{\xi}}  \phi(\bm{\xi}) \right)^2   \right]                            T_{1}\left( \frac{k\bm{\xi} }{f_{1} - \Delta f} + \frac{k\bm{\rho} }{f_{2}}\right) \nonumber \\ && \times
\int d\bm{\eta} E_{0i}^{\ast} e^{-i\phi(\bm{\eta})}  \left[1 +   \frac{\Delta f}{2k} \nabla_{\bm{\eta}}^2  \phi(\bm{\eta})      - \frac{i\Delta f}{2k}\left( \nabla_{\bm{\eta}}  \phi(\bm{\eta}) \right)^2   \right]                            T_{1}\left( \frac{k\bm{\eta} }{f_{1} - \Delta f} + \frac{k\bm{\rho} }{f_{2}}\right).
\end{eqnarray}
Note that it is not necessary to take the conjugate of the function $T_{1}$ because it is a real quantity. By multiplying the three terms in each integral for their conjugates we obtain nine terms. By neglecting all quadratic powers of  $\Delta f$ and considering $E_{0i}E_{0i}^{\ast}$ a constant factor we obtain the final expression for intensity
\begin{eqnarray}
I(\bm{\rho})= \int d\bm{\xi} \int d\bm {\eta} e^{i[\phi(\bm{\xi}) -\phi(\bm{\eta}) ]}  T_{1}\left( \frac{k\bm{\xi} }{f_{1}-\Delta f} + \frac{k\bm{\rho} }{f_{2}}\right)      T_{1}\left( \frac{k\bm{\eta} }{f_{1}-\Delta f} + \frac{k\bm{\rho} }{f_{2}}\right)  \nonumber \\  \times
\left\{ 1  +  \frac{\Delta f}{2k}\left[     \left(  \nabla_{\bm{\xi}}^2\phi(\bm{\xi})   +     \nabla_{\bm{\eta}}^2\phi(\bm{\eta}) \right)
+i\left( \left( \nabla_{\bm{\xi}}\phi(\bm{\xi}) \right)^2  -  \left(\nabla_{\bm{\eta}}\phi(\bm{\eta})\right)^2                             \right)\right] \right\}.
\end{eqnarray}
This is a real function. The approximation $\frac{1}{f_{1} - \Delta f } \approx \frac{1}{f_{1}} + \frac{\Delta f}{f_{1}^2}$ used in the quadratic phase factors above implies that we may replace $f_{1} - \Delta f$ by $f_{1} $ in the denominator of the argument of $T_{1}$, which gives
\begin{eqnarray}
I(\bm{\rho})= \int d\bm{\xi} \int d\bm {\eta} e^{i[\phi(\bm{\xi}) -\phi(\bm{\eta}) ]}  T_{1}\left( \frac{k\bm{\xi} }{f_{1}} + \frac{k\bm{\rho} }{f_{2}}\right)      T_{1}\left( \frac{k\bm{\eta} }{f_{1}} + \frac{k\bm{\rho} }{f_{2}}\right) \label{eq final} \nonumber \\ \times
\left\{ 1  +  \frac{\Delta f}{2k}\left[     \left(  \nabla_{\bm{\xi}}^2\phi(\bm{\xi})   +     \nabla_{\bm{\eta}}^2\phi(\bm{\eta}) \right)
+i\left( \left( \nabla_{\bm{\xi}}\phi(\bm{\xi}) \right)^2  -  \left(\nabla_{\bm{\eta}}\phi(\bm{\eta})\right)^2                             \right)\right] \right\}.
\end{eqnarray}
When the size of the lens aperture tends to infinite we have $T_{1}\left( \frac{k\bm{\alpha} }{f_{1}} + \frac{k\bm{\rho} }{f_{2}}\right) \rightarrow \delta\left( \frac{k\bm{\alpha} }{f_{1}} + \frac{k\bm{\rho} }{f_{2}}\right)$, where $\alpha=\xi , \,\eta$ .
In this case, the delta functions filter the integrals, so that
\begin{eqnarray}
I(\bm{\rho})= 1 + \frac{\Delta f}{k}\nabla_{\rho'}^2 \left[ \phi\left( \bm{\rho}'  \right)\right],
\label{eq z}
\end{eqnarray}
which reproduces Eq. (\ref{eq bira}) as a particular case. Also, $\bm{\rho}' = -\frac{f_{1}}{f_{2}} \bm{\rho}$ and the subscript $\rho'$ indicates that the laplacian is calculated with respect to this variable.

Equation (\ref{eq final}) is the main theoretical result of this paper. It is a generalization of the previous treatment given for DM in Ref. \cite{Agero03}. Note that the gradient and the phase itself are now present in the expression for the intensity in the image plane. So, an object that introduces only linear changes in the phase of the field can be detected by DM with the use of a small objective aperture. This is not possible when an infinite objective aperture is considered.

\begin{figure}[t]
\centering
\includegraphics[width=0.7\textwidth]{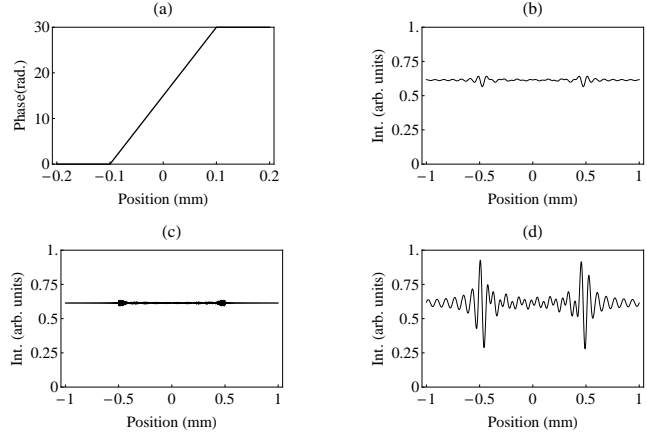}
\caption{(a) Object complex transmission function. (b) The intensity at the image plane when $2L= 10$\,mm and $\Delta f = 0$. (c) The intensity at the image plane of the same phase object, but now considering $\Delta f = 2$\,mm and $2L= 200$\,mm. (d) The intensity at the image plane for $2L= 10$\,mm, keeping $\Delta f = 2$\,mm (See the main text for details).} \label{grad}
\end{figure}

Figures \ref{grad}(b)-(d) show the intensities at the image plane for a pure phase object that introduces a linear phase change into the field, as shown in Fig. \ref{grad}(a). In this case, Eq. (\ref{eq bira}) predicts only an uniform backlight. These curves were calculated via Eq. (\ref{eq final}) with the parameters $f_{1}=100$\,mm, $f_{2}=500$\,mm, and $826$\,mm of wavelength. Figure \ref{grad}(b) shows the intensity at the image plane when $2L= 10$\,mm and $\Delta f = 0$. The oscillations around the points $\rho = \pm 0.5$\,mm at the image plane are related to the points where there are abrupt phase changes at the object plane, considering the magnification of the imaging system, $M=-f_{2}/f_{1}=-5$. In Fig. \ref{grad}(c), we have the intensity at the image plane of the same phase object, but now considering $\Delta f = 2$\,mm. It was also assumed an objective lens aperture of $2L= 200$\,mm. The idea of using a large objective aperture is to simulate an infinite lens. Note that the values of $f_{1}$ and $\Delta f$ are compatible with the approximations $\frac{1}{f_{1} - \Delta f } \approx \frac{1}{f_{1}} + \frac{\Delta f}{f_{1}^2}$ in the quadratic phase factors and $f_{1} - \Delta f \approx f_{1} $ in the denominator of the argument of $T_{1}$. The contrast of the diffracted images tends to zero when the size $2L$ of the objective lens tends to infinite. This shows that a small objective aperture can be more effective than defocusing the system in some situations. Finally, Fig. \ref{grad}(d) shows the intensity for a lens aperture of $2L= 10$\,mm, keeping $\Delta f = 2$\,mm. For this case, the intensity has more contrast in comparison with the last ones due to the joint action of the finite size of the objective lens aperture and the defocus of the objective lens.

\section{Experiment}

\begin{figure}[ht]
	\centering
		\includegraphics[width=0.5\textwidth]{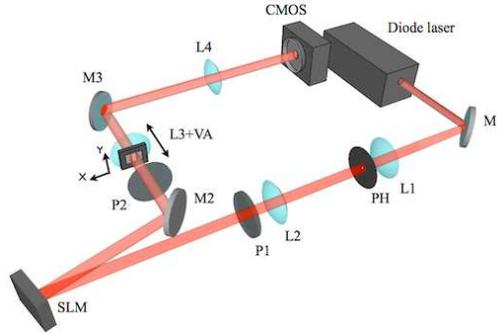}
		\caption{Schematic draw of the experimental setup. $L_{1}$ is a $25$\,mm focal length lens. The focal length of $L_{2}$ is $200$\,mm. The focal lengths of $L_{3}$ and $L_{4}$ are both $300$\,mm. $L_{3}$ is mounted on a motorized linear stage. PH is a $100$\,$\mu$m diameter pinhole. $P_{i}$ ($i=1,2$) are polarizers. SLM is the LCD of the phase spatial light modulator. $M_{i} (i=1,2,3)$ are mirrors. Diode laser is a $633$\,nm laser. CMOS is a camera. All lenses have $1''$ in diameter.  }\label{fig2}
\end{figure}

\begin{figure}[ht]
	\centering
		\includegraphics[width=0.4\textwidth]{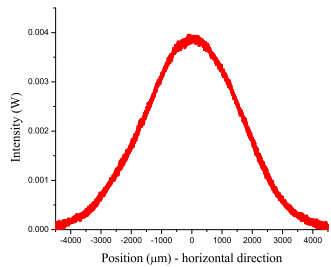}
		\caption{Beam profile (at the horizontal direction) of the laser after crossing the telescope. The data was obtained with a scanning slit optical beam profiler.}\label{figProf}
\end{figure}

We performed an experiment to verify some predictions of our theoretical generalization of DM, presented in the last section. More specifically, we experimentally study the dependence of the image contrast with the aperture of the objective lens of a microscope system. The experimental setup is showed in Fig. \ref{fig2}. A single mode diode laser operating in TEM$_{00}$ mode with $\sim 0.5$\,mm FWHM and at $633$\,nm wavelength was used as the source of illumination. This beam crosses a telescope lens system with a spatial filter to improve its quality and the beam spot size. After the telescope, the beam is well collimated and magnified by a factor $8$. We used lenses L$1$ and L$2$ ($1''$ in diameter) that have $25$\,mm and $200$\,mm of focal lengths, respectively. We also used a pinhole (PH) of $100$\,$\mu$m of diameter that was placed between the lenses. This allowed us to obtain the spatial beam profile shown in Fig. \ref{figProf}. It shows that for illuminated objects smaller than $1$\,mm, it is reasonable to assume that the incident field is a plane wave. The beam is sent to a reflective spatial light modulator (SLM), which is composed of two polarizers (P$1$ and P$2$) and a twisted nematic liquid crystal display (LCD), model Holoeye $1080$P. The SLM is adjusted for phase modulation only \cite{Lima1}.

\begin{figure}[ht]
	\centering
		\includegraphics[angle=-90,width=0.7\textwidth]{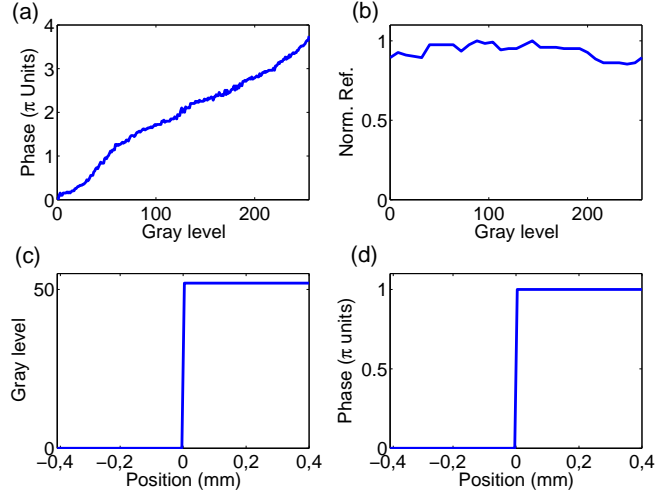}
		\caption{(a) Calibration curve of the SLM for phase modulation at $633$\,nm wavelength. (b) Amplitude modulation versus the LCD gray level. (c) Modulation used and (d) the corresponding generated object at the SLM.}\label{fig3}
\end{figure}

Figure \ref{fig3}(a) shows the curve of the phase modulation versus the LCD gray level for our SLM at $633$\,nm wavelength. Using the SLM at the phase modulation configuration, we tested if there was still any amplitude modulation in the incident field. The amplitude as a function of the LCD gray level is shown in Fig. \ref{fig3}(b). This curve ensures that the equipment is modulating the phase of the field with a fluctuation of less than $10$\% in the amplitude modulation. After ensuring the capability to generate a reasonable pure phase object, we generate at the SLM an object that introduces a phase discontinuity at the incident field equal to $\pi$ [See Fig. \ref{fig3}(c)], as shown in Fig. \ref{fig3}(d). For simplicity, we considered a one-dimensional phase object. It simulates the effect of the edge of a transparent object over a plane wave field, supposing that this field has crossed the object in a perpendicular direction to its plane face \cite{Faust55,Stagaman}. In fact, since a pixel of this equipment has $8$\,$\mu$m of linear dimension, and also because the SLM is based on a reflective LCD, we can guarantee that the phase discontinuity occurs in a very small zone over the horizontal direction. To perform a $\pi$ phase change, we have chosen $0$ for the LCD gray level at the left side of the discontinuity and $52$ for the LCD gray level at the right side. The object was generated at the SLM such that the discontinuity was located at the center of the incident pump beam.

After being reflected by the LCD, the light follows to a microscope system, which is prepared with the lenses L$3$ (objective) and L$4$ (eyepiece), as can be seen in Fig. \ref{fig2}. They have $1''$ of diameter and equal focal lengths of $300$\,mm. In this system, the LCD is placed at the focal distance of L$3$, L$4$ is placed at a distance of $600$\,mm from L$3$, and the camera is positioned at $300$\,mm after L$4$ at the image plane. The rectangular variable aperture (VA) is placed right in front of the lens L$3$ in order to control the dimension of the entrance pupil of the system (see Fig. \ref{fig2}). It has a fixed length in the vertical direction and variable length in the horizontal direction. Its horizontal length is controllable by a screw with a $0.005''$ of precision. The VA together with L$3$ constitute an effective objective lens with a controllable horizontal linear aperture. To control the defocus, the system VA$+$L$3$ is placed over a motorized stage that allows for the control of its longitudinal position with a precision of $100$\,$\mu$m. The camera used to obtain the images is a CMOS, model EO-$1312$M $1/2''$ Monochrome USB Lite Edition, from Edmond Optics.

\begin{figure}[ht]
	\centering
		\includegraphics[angle=-90,width=0.7\textwidth]{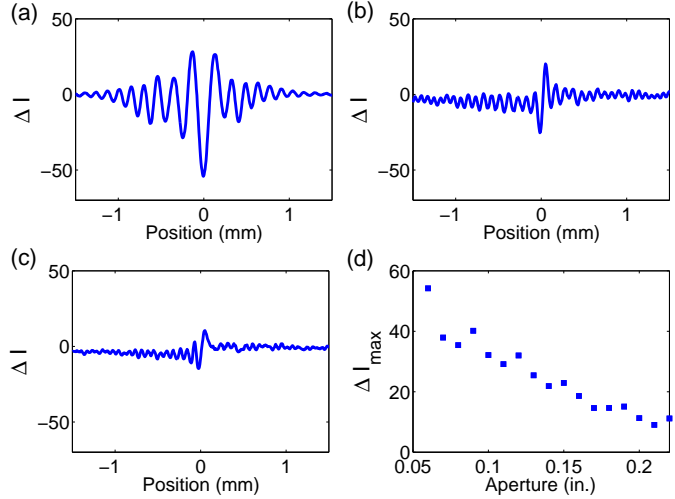}
		\caption{(a), (b), and (c) show the intensities of the noise-subtracted images versus horizontal position in the plane of the camera for apertures equal to $0.0600''$, $0.130''$, and $0.190''$, respectively. (d) shows $\Delta I_{max}$ versus the linear size of the objective lens aperture.}\label{fig4}
\end{figure}

\begin{figure}[ht]
	\centering
		\includegraphics[width=0.6\textwidth]{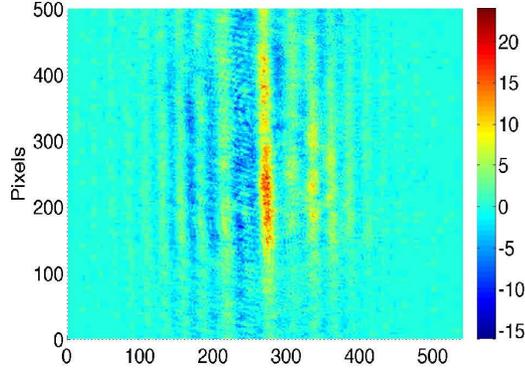}
		\caption{The noise-subtracted image for the horizontal aperture of VA equal to $0.100''$. Each pixel of the CMOS has $5.2$\,$\mu$m.}\label{figImage}
\end{figure}

Figure \ref{fig4} shows the obtained experimental results. The interference patterns were recorded with $\Delta f=0$, in order to verify the effects associated with the variation of the size of the entrance pupil only. Figures \ref{fig4}(a), \ref{fig4}(b), and \ref{fig4}(c) show the intensities of the noise-subtracted images, while considering VA equal to $0.0600''$, $0.130''$, and $0.190''$, respectively. Given an aperture, two images were measured. They correspond to the case where the object was being or not generated in the SLM. When SLM generates the object, the image has contributions from the noisy background and from the phase modulation performed by the device. This means that it is necessary to subtract the noise from the image to proper take into account the effect of the entrance pupil in the Becke line. To do this, we recorded the contribution to the image corresponding to the background noise. This is done by generating an uniform global phase object at the SLM, such that it works as a conventional mirror. This procedure reveals all non-desired modulations at the image plane. Then, we subtract this noisy image from the signal recorded before. An example of a noise-subtracted image can be seen in Fig. \ref{figImage}. Note that the subtracted intensity ($\Delta I$) can be negative since the intensity of the noise in a pixel can be greater than the diffracted image when the SLM is generating the object.

In order to evaluate $\Delta I$, we performed a statistical analysis of each noise-subtracted image obtained. Each image was treated as a matrix, where each element was associated with a physical pixel of the camera. The corresponding $\Delta I$ for a certain horizontal position at the image plane was calculated by taking the mean value of the intensities of each pixel in a column. Each column was associated with a certain horizontal position. As one can see from Figs. \ref{fig4}(a), \ref{fig4}(b) and \ref{fig4}(c), the contrast increases when the linear horizontal dimension of VA is reduced. Fig. \ref{fig4}(d) shows the curve for $\Delta I_{max}$ as a function of the apertures measured. It is possible to associate $\Delta I_{max}$ directly with the contrast of the image observed \cite{Agero03}, where larger $\Delta I_{max}$ implies higher contrast.

\section{Conclusion}
We have studied theoretically and experimentally the defocusing microscopy technique by taking into account the finite size of the objective lens aperture. We concluded that by using smaller objective lens aperture one can detect linear phase changes caused by phase objects, that could not be detected when infinite (very large) lens are considered (used). This improvement may lead to new applications for DM, such as the observation of ``hidden'' structures in cells.

We also observed that small lens apertures generates diffracted images with higher contrast than lenses with infinite aperture diameters.. This type of manipulation can be combined with defocusing to improve the efficiency of the DM technique. It is important to note that DM is a low resource consuming technique that can easily be applied in any microscope. Thus, the work presented here will be relevant in the study of any phase object in different areas of science.

\appendix
\section{\textbf{Appendix: Microscopic System}}

Using the Fresnel propagators \cite{Goodman}, we found that the electric field at the image plane for the apparatus shown in figure \ref{fig1}(b) is given by

\begin{eqnarray}
E(\bm{\rho}) = \int
d\bm{\xi} E_{0} \left( \bm{\xi} \right)   \int d\bm{\alpha} \left| A_{1}(\bm{\alpha})\right | e^{\frac{-ik \alpha ^{2}}{2f_{1}}}    e^{\frac{ik |\bm{\alpha} - \bm{\xi|}^{2}}{2f_{1}}}   \nonumber         \\
   \int d\bm{\beta}\left| A_{2}(\bm{\beta}) \right | e^{\frac{-ik \beta ^{2}}{2f_{2}}}    e^{\frac{ik |\bm{\beta} - \bm{\alpha|}^{2}}{2(f_{1} +f_{2})}}    e^{\frac{ik |\bm{\rho} - \bm{\beta|}^{2}}{2f_{2}}} ,
\label{ap a}
\end{eqnarray}
 where  $\left| A_{1}(\bm{\alpha})\right|$  and $\left| A_{2}(\bm{\beta})\right| $ are the modulus of transmission function of objective and ocular lens respectively. The integrations in $\bm{\xi}$,  $\bm{\alpha}$ and  $\bm{\beta}$  implement the propagations from the object to the objective, from the objective to the ocular and from the ocular to the image plane respectively. The Eq. (\ref{ap a}) can be written as
\begin{eqnarray}
E(\bm{\rho}) =   e^{\frac{ik \rho ^{2}}{2f_{2}}} \int
d\bm{\xi} E_{0} \left( \bm{\xi} \right)    e^{\frac{ik \xi ^{2}}{2f_{1}}}    g(\bm{\xi} , \bm{\rho})
\label{yyyy}
\end{eqnarray}
where
\begin{eqnarray}
g(\bm{\xi} , \bm{\rho}) =    \int d\bm{\alpha}   \int d\bm{\beta}  \left| A_{1}(\bm{\alpha})\right |   \left| A_{2}(\bm{\beta}) \right|
 e^{\frac{ik |\bm{\beta} - \bm{\alpha|}^{2}}{2(f_{1} + f_{2})}}
e^{\frac{-ik \bm{\alpha} \cdot \bm{\xi}}{f_{1}}}   e^{\frac{-ik \bm{\rho} \cdot \bm{\beta}}{f_{2}}}.
\end{eqnarray}
Let us define
\begin{eqnarray}
 \frac{k \bm{\xi}}{f_{1}}= \bm{x}_{1},  \ \ \ \ \ \ \ \ \ \ \ \   \frac{k \bm{\rho}}{f_{2}}= \bm{x}_{2}, \ \ \ \ \ \ \ \ \ \ \ \   \frac{k}{2(f_{1} + f_{2})} = y .
\end{eqnarray}
In this way we have
\begin{eqnarray}
g(\bm{x}_{1} , \bm{x}_{2}) =  \texttt{FF}\left\{  \left| A_{1}(\bm{\alpha})\right |   \left| A_{2}(\bm{\beta}) \right| \ \
e^{iy |\bm{\beta} - \bm{\alpha|}^{2}}  \right\},
\end{eqnarray}
where the symbol $\texttt{FF}$ mean the double Fourier transform with respect to the variable par $(\bm{\alpha}, \bm{\beta})$.
Using the convolution theorem \cite{Goodman} we have
\begin{eqnarray}
g(\bm{x}_{1} , \bm{x}_{2}) =  \texttt{FF}\left\{  \left| A_{1}(\bm{\alpha})\right |   \left| A_{2}(\bm{\beta}) \right| \right\} \ast
\   \texttt{FF} \left\{e^{iy |\bm{\beta} - \bm{\alpha|}^{2}}  \right\},
\end{eqnarray}
where the symbol $\ast$ means the convolution. The double Fourier transform assumes the factorable form
\begin{eqnarray}
\texttt{FF}\left\{  \left| A_{1}(\bm{\alpha})\right |   \left| A_{2}(\bm{\beta}) \right| \right\} =
\texttt{F} \left\{ \left| A_{1}(\bm{\alpha})\right |  \right\}  \texttt{F} \left\{ \left| A_{2}(\bm{\beta})\right |  \right\} = T_{1}(\bm{x}_{1})T_{2}(\bm{x}_{2}).
\end{eqnarray}
Let us define the function $h(\bm{x}_{1} , \bm{x}_{2})$ as
\begin{eqnarray}
h(\bm{x}_{1} , \bm{x}_{2}) =    \texttt{FF} \left\{e^{iy |\bm{\beta} - \bm{\alpha|}^{2}}  \right\}
\end{eqnarray}
so that
\begin{eqnarray}
g(\bm{x}_{1} , \bm{x}_{2}) = T_{1}(\bm{x}_{1})T_{2}(\bm{x}_{2}) \ast  h(\bm{x}_{1} , \bm{x}_{2})
\label{ap g}.
\end{eqnarray}
After some calculations it is possible to show that
\begin{eqnarray}
h(\bm{x}_{1} , \bm{x}_{2}) = \delta (\bm{x}_{1} + \bm{x}_{2}) e^{-i\frac{|\bm{x}_{2} - \bm{x}_{1}|^2}{16y}}.
\label{ap h}
\end{eqnarray}
In order to demonstrate the above equation we suggest the following change of variables: $\bm{\beta} + \bm{\alpha}  = 2\bm{u} $ and $\bm{\beta} - \bm{\alpha}  = 2\bm{v} $. These changes will eliminate the cross terms so that the expression will become the product of two independents integrals. Inserting Eq. (\ref{ap h}) in Eq. (\ref{ap g}) we have
\begin{eqnarray}
g(\bm{x}_{1} , \bm{x}_{2}) = T_{1}(\bm{x}_{1})T_{2}(\bm{x}_{2}) \ast \delta (\bm{x}_{1} + \bm{x}_{2}) e^{-i\frac{|\bm{x}_{2} - \bm{x}_{1}|^2}{16y}},
\end{eqnarray}
that is,
\begin{eqnarray}
g(\bm{x}_{1} , \bm{x}_{2}) = \int d\bm{x'}_{1} \int d\bm{x'}_{2}    T_{1}(\bm{x}_{1} - \bm{x'}_{1})T_{2}(\bm{x}_{2} - \bm{x'}_{2})  \delta (\bm{x'}_{1} + \bm{x'}_{2}) e^{-i\frac{|\bm{x'}_{2} - \bm{x'}_{1}|^2}{16y}}.
\end{eqnarray}
Solving the above integral and going back to the original variables we obtain
\begin{eqnarray}
g(\bm{\xi} , \bm{\rho}) = \int d\bm{v} T_{1}\left(\frac{k \bm{\xi}}{f_{1}} - \bm{v}  \right)  T_{2}\left(\frac{k \bm{\rho}}{f_{2}} + \bm{v}  \right)
e^{\frac{-i(f_{1} +  f_{2})}{2k} v^2}.
\end{eqnarray}
This is the optical transfer function (OTF) of the microscope system, which takes into account the finite size of both objective and ocular apertures.
Assuming an infinite ocular aperture, so that $T_{2}\left(\frac{k \bm{\rho}}{f_{2}} + \bm{v}  \right) = \delta \left(\frac{k \bm{\rho}}{f_{2}} + \bm{v}  \right)$ we conclude that
\begin{eqnarray}
g(\bm{\xi} , \bm{\rho}) =  T_{1}\left(\frac{k \bm{\xi}}{f_{1}} + \frac{k \bm{\rho}}{f_{2}} \right)
e^{\frac{-ik(f_{1} +  f_{2})}{2f_{2}^2} \rho^2}.
\label{OTF}
\end{eqnarray}
Inserting the Eq. (\ref{OTF}) in Eq. (\ref{yyyy}), we find
\begin{eqnarray}
E(\bm{\rho}) =   const. \ \ e^{\frac{ik \rho ^{2}}{2f_{2}}}   e^{\frac{-ik(f_{1} +  f_{2})}{2f_{2}^2} \rho^2}  \int
d\bm{\xi} E_{0} \left( \bm{\xi} \right)    e^{\frac{ik \xi ^{2}}{2f_{1}}}  T_{1}\left(\frac{k \bm{\xi}}{f_{1}} + \frac{k \bm{\rho}}{f_{2}} \right).
\label{telescopico final}
\end{eqnarray}
According \cite{Goodman}, the quadratic phase factor $e^{\frac{ik \xi ^{2}}{2f_{1}}}$  in Eq. (\ref{telescopico final}) causes  ``an unacceptably large image blur''. In order to eliminate it, we consider one of the following two restrictions. First, we can assume that the object is illuminated  by a spherical wave that is converging towards the point where the optical axis pierces the lens, that is, the incident field contains the term $e^{\frac{-ik \xi ^{2}}{2f_{1}}}$  which cancels the quadratic exponential \cite{Goodman}. Second, we can assume that the size of object is no greater than about $1/4$ the size of the lens aperture \cite{Goodman,Tichenor}. After assuming one of these restrictions, we have finally the expression
\begin{eqnarray}
E(\bm{\rho}) =   const. \ \ e^{\frac{ik \rho ^{2}}{2f_{2}}}   e^{\frac{-ik(f_{1} +  f_{2})}{2f_{2}^2} \rho^2}  \int
d\bm{\xi} E_{0} \left( \bm{\xi} \right)    T_{1}\left(\frac{k \bm{\xi}}{f_{1}} + \frac{k \bm{\rho}}{f_{2}} \right).
\label{}
\end{eqnarray}
The phase factor outside the integral vanishes only when the size of both (ocular and objective) lenses tends to infinity.

\section*{Acknowledgements}

This work was supported by Grants Milenio P$10$-$030$-F and PFB $08024$.  S. P\'{a}dua acknowledges the support of CNPq, FAPEMIG and Instituto Nacional de Ci\^{e}ncia e Tecnologia - Informa\c{c}\'{a}o Qu\^{a}ntica. Ivan F. Santos acknowledges the support of FAPEMIG.

\end{document}